\documentclass[12pt,preprint,natbib]{emulateapj}
\def \HST{{\emph{HST}}}
\def \ha{{H$\alpha$}}
\def \hb{{H$\beta$}}
\def \hahb{{H$\alpha$/H$\beta$}}
\def \Spitzer{{\emph{Spitzer}}}
\def \tf {{24 $\mu$m}}
\def \um {{$\mu$m}}
\def \boot {{Bo\"otes}}
\def \sm {{M$_\odot$}}
\def \yr {{yr$^{-1}$}}

\begin{document}

\title{Black Hole Masses and Star Formation Rates of $z >1$ Dust Obscured
Galaxies (DOGs):\\ Results from Keck OSIRIS Integral Field Spectroscopy}

\author{J. Melbourne \altaffilmark{1}, Chien Y. Peng \altaffilmark{2},
B. T. Soifer \altaffilmark{1,3}, Tanya Urrutia \altaffilmark{3}, Vandana Desai \altaffilmark{3}, L. Armus
\altaffilmark{3}, R. S. Bussmann \altaffilmark{4}, Arjun Dey \altaffilmark{5},
 K. Matthews \altaffilmark{1}}

\altaffiltext{1}{Caltech Optical Observatories, Division of Physics,
Mathematics and Astronomy, Mail Stop 301-17, California Institute of
Technology, Pasadena, CA 91125, jmel@caltech.edu, bts@ipac.caltech.edu,
kym@caltech.edu}

\altaffiltext{2}{Herzberg Institute of Astrophysics, National Research Council
of Canada, 5071 West Saanich Road, Victoria, British Columbia V9E 2E7, Canada,
cyp@nrc-cnrc.gc.ca}

\altaffiltext{3}{Spitzer Science Center, Mail Stop 314-6, California Institute
of Technology, Pasadena, CA 91125, bts@ipac.caltech.edu, turrutia@ipac.caltech.edu, desai@ipac.caltech.edu,
lee@ipac.caltech.edu }

\altaffiltext{4}{University of Arizona, LBT Observatory, 933 N. Cherry Ave.,
Tuscon, AZ 85721-0065, rsbussmann@as.arizona.edu}

\altaffiltext{5}{National Optical Astronomy Observatory, P.O. Box 26732,
Tucson, AZ 85726-6732, dey@noao.edu, jannuzi@noao.edu}

\begin{abstract}

We have obtained high spatial resolution Keck OSIRIS integral field
spectroscopy of four $z\sim1.5$ ultra-luminous infrared galaxies that exhibit
broad \ha\ emission lines indicative of strong AGN activity.  The observations
were made with the Keck laser guide star adaptive optics system giving a
spatial resolution of $0.1\arcsec$ or $<1$ kpc at these redshifts.  These
high spatial resolution observations help to spatially separate the extended
narrow-line regions --- possibly powered by star formation --- from the nuclear regions, which
may be powered by both star formation and AGN activity.  There is no evidence for 
extended, rotating gas disks in these four galaxies.  Assuming dust
correction factors as high as $A(H\alpha)=4.8$ mag, the observations suggest
lower limits on the black hole masses of $(1 - 9)\times10^8 $ \sm, and star formation rates
$<100$ \sm\ \yr.  The black hole masses and star formation rates of the sample galaxies appear low in comparison to other high-$z$ galaxies with similar host luminosities.  We explore possible explanations for these observations including, host galaxy fading, black hole growth, and the shut down of star formation.

 

\end{abstract}


\section{Introduction}

\Spitzer\ \tf\ imaging in the 9 square degree NOAO Deep Wide Field Survey \citep[NDWFS;][]{JannuziDey99} of \boot\ has revealed large samples of high redshift ultra-luminous infrared galaxies (ULIRGs; $L_{IR} =10^{12-13}$).   A simple optical-to-IR color cut of  $R-[24] > 14$ Vega mag, i.e. $f_{\nu}$(\tf)$ / f_{\nu} (R) \ga 1000$, selects for an extreme class of ULIRGs at redshifts of  $< z > \ =1.99$, $\sigma_z =0.5$ \citep{Dey08}. ULIRGs selected this way are known as Dust Obscured Galaxies (DOGs), and are typically redder than low redshift ULIRGs \citep{Dey08}. The extreme luminosities and colors of DOGs suggest the presence of AGN accretion and/or intense star-formation heavily obscured by dust at rest-frame optical and UV wavelengths. 

The rest-frame optical-to-mid IR (MIR) spectral energy distributions (SEDs) of the DOGs show two classes: (1) power law sources with SEDs that rise smoothly into the MIR;  and (2) so called ``Bump" sources, with a rest-frame 1.6 $\mu$m peak in their SED thought to be produced by the photospheres of cool stars.  The power-law DOGs have been shown to exhibit strong AGN characteristics including broad-H$\alpha$ \citep{Brand07}, and a lack of PAH emission in MIR spectra from \Spitzer\ \citep{Houck05}.  Meanwhile MIR spectra of bump DOGs tend to show PAH emission \citep{Desai08}, a good indication of vigorous ongoing star formation.  These two classes of DOGs appear to differentiate with luminosity, with the more luminous sources (e.g. $F_{24} > 0.75$ mJy) having a higher fraction of power-law SEDs likely to harbor obscured AGN \citep{Magliocchetti07,Dey08}.

The  space densities and clustering strength of the DOGs are similar to sub-mm galaxies \citep[SMGs;][]{Blain04,Magliocchetti07,Brodwin08}, which are thought to have merger driven star formation rates as high as 1000 $M_{\odot}$ yr$^{-1}$.  The clustering strengths are also similar to massive elliptical galaxies in the local universe, suggesting a possible connection between these three different galaxy classes.

Detailed merger simulations of massive gas rich galaxies have been shown to reproduce the selection criteria of both SMGs and DOGs \citep{Narayanan09a}.  In these simulations, the SMGs are typically classified as bump sources, and the SMG phase is followed by a hotter, post-merger, AGN dominated phase where the galaxy is transformed into a power-law DOG \citep{Narayanan09a}.  Currently, mergers are the only model shown to produce the extreme colors and luminosities of DOGs.

Because of their potential link to gas rich mergers, it was surprising to find that the rest-frame optical morphologies of the more luminous DOGs are typically smooth with little sign of an ongoing merger \citep[as evidenced by double nuclei,][]{Melbourne08b,Bussmann09,Melbourne09}.  In fact, the bulk of DOGs with high spatial resolution imaging (from \HST\ or Keck adaptive optics) show disk-like or elliptical like morphologies, rather than double nuclei or tidal tails.  However, their physical half-light sizes are small in comparison with LIRGs at $z\sim1$, which also favor disk-like profiles; the more luminous DOGs have typical half-light sizes that are a factor of 2 (or more) smaller than sample of $z\sim1$ LIRGs \citep{Melbourne08a,Melbourne08b}.  The unusual sizes of the DOGs are consistent with the post-merger products of gas rich mergers in the local universe \citep[e.g.][]{Rothberg06}.  

If DOGs have undergone recent merger activity, evidence for the merger should be imprinted on the kinematics of the gas in the galaxy.  While they exhibit exponential disk profiles, the gas in the DOGs may not necessarily be undergoing ordered rotation.  With its high spatial and spectral resolution, the Keck OSIRIS integral field spectrograph (IFS) is a ideal tool for measuring gas kinematics in high-$z$ galaxies.  OSIRIS is an adaptive optics fed instrument with a diffraction limited angular resolution of $0.05\arcsec$ at $K$-band (2.2 \um).  OSIRIS produces a spectrum at every spatial resolution element over a field of view of $2-3$ arcsecs.  With a spectral resolution of $R\sim3000$, OSIRIS can resolve kinematic signatures with FWHM of 100 km s$^{-1}$.  Targeting the H$\alpha$ emission line redshifted into the near-infrared (NIR), OSIRIS has been used to determine the kinematics of $z=1-3$ galaxy samples \citep[e.g.][]{Wright07, Law09, Wright09}.

Because of its high spatial resolution, OSIRIS also offers the possibility of spatially differentiating the extended star forming region from the central regions which may contain both star formation and AGN  activity.  This capability allows for much cleaner determinations of the relative strengths of these two power sources, compared with seeing limited observations.  

This paper presents OSIRIS observations of four DOGs at $z>1$.   Each DOG was targeted at the wavelength of the H$\alpha$ line redshifted into the NIR.  We quantify the flux contributions from broad and narrow \ha\ emission-line regions, and use the data to place constraints on supermassive black hole masses, star formation rates (SFRs), and metallicities of the DOGs.  We also examine the kinematics of each system.

This paper is organized as follows.  The sample selection and OSIRIS observations are described in Section 2.  Section 3 details the \ha\ line measurements.  Properties of the sample derived from spectra, including the black hole masses and star formation rates, are presented in Section 4.  The results are discussed in Section 5, and we compare the DOGs to other high-$z$ and local AGN hosts. Conclusions are provided in Section 6.  Throughout we assume a canonical $\Lambda$ Cold Dark Matter Universe with $\Omega_M=0.3$, $\Omega_{\Lambda}=0.7$, and $H_0=70$ km s$^{-1}$ Mpc$^{-1}$.

\section{OSIRIS Integral Field Spectroscopy}

\begin{deluxetable*}{ccccccccc}
\tabletypesize{\small}
\tablecaption{DOG Properties Summary \label{tab:prop}}
\tablehead{
\colhead{Galaxy} &\colhead{Object Name}  & \colhead{redshift}  & \colhead{f$_{\nu}$(24)} & \colhead{R}  \\
&    & &   \colhead{[mJy]}  & \colhead{[Vega mag]}}
\startdata
NDWFS\_J143027.2+344008 & DOG1 & 1.370	 &  1.169     &	  24.76 \\
NDWFS\_J143335.6+354243 & DOG2 & 1.300	 &  5.577     &	  21.85 \\
NDWFS\_J143400.3+335714 & DOG3 & 1.684	 &  1.754     &	  23.89 \\
NDWFS\_J143424.5+334542 & DOG4 & 2.260	 &  0.860     &	  25.60 \\
\enddata			 
\end{deluxetable*}

We used the Keck laser guide star adaptive optics (LGSAO) system and the OSIRIS IFS to observe four $z>1$ ULIRGs. These galaxies were selected from optical and MIR photometry of the NDWFS of \boot\ to have $R - [24] > 14$ Vega mag, and are thus DOGs.  The sample selection also required that each DOG have a strong \ha\ detection in seeing limited NIR spectroscopy.  These spectra provided not only a redshift, but also an \ha\ flux.  We selected those DOGs with the strongest \ha\ detections that were sufficiently near to an AO tip-tilt guide star, to provide excellent AO performance.   

\begin{figure*}
\centering
\plottwo{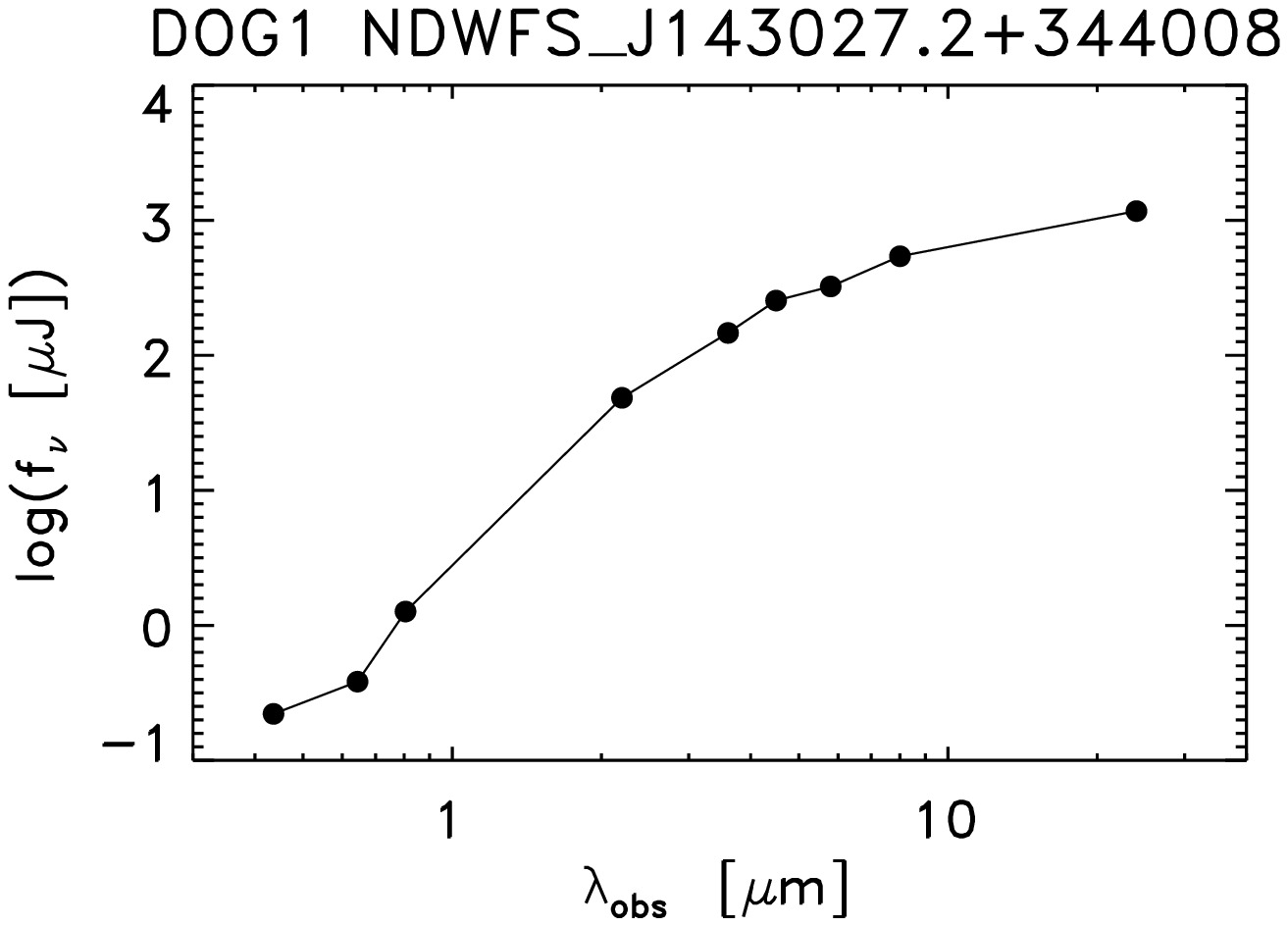}{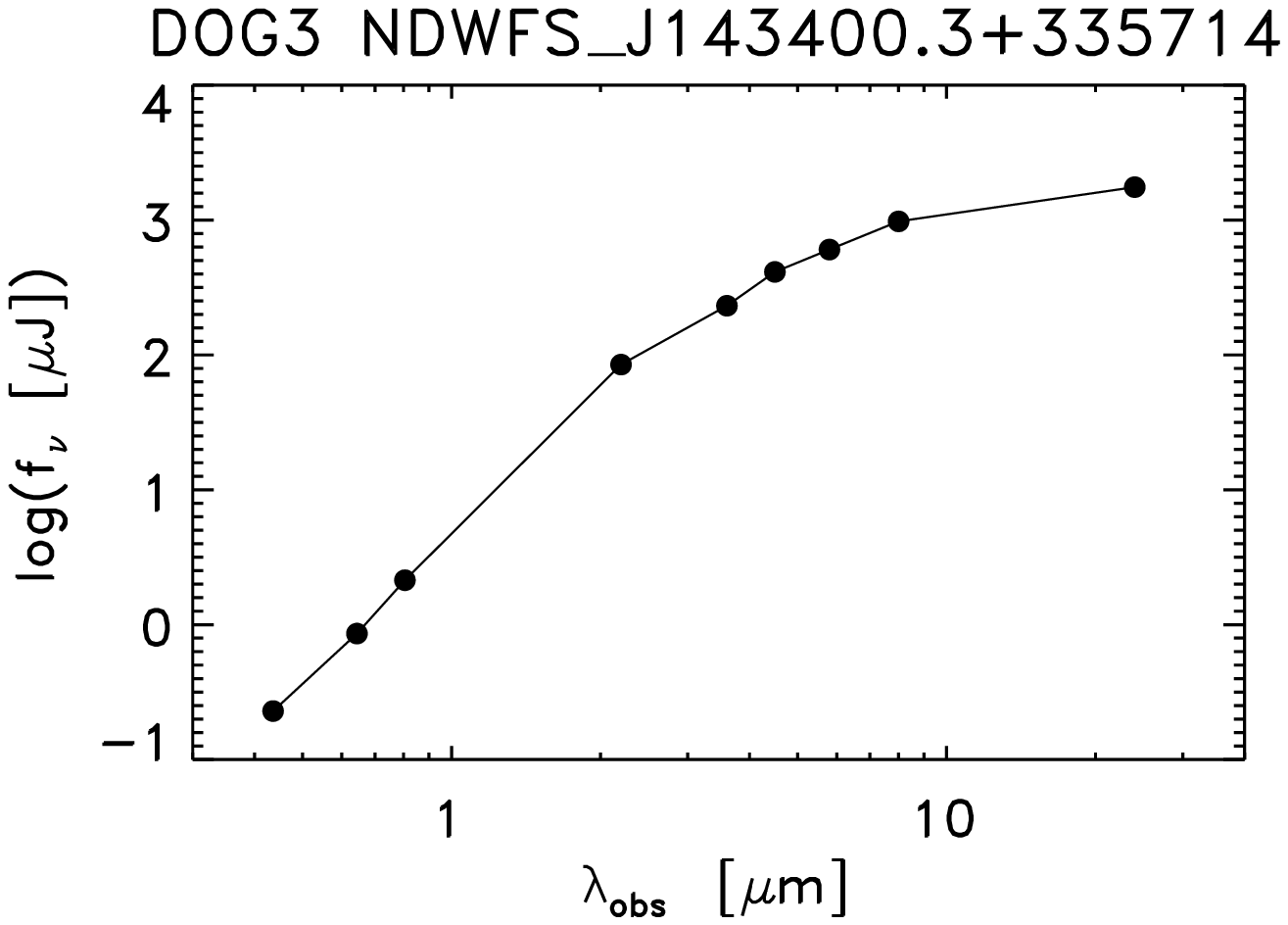}
\plottwo{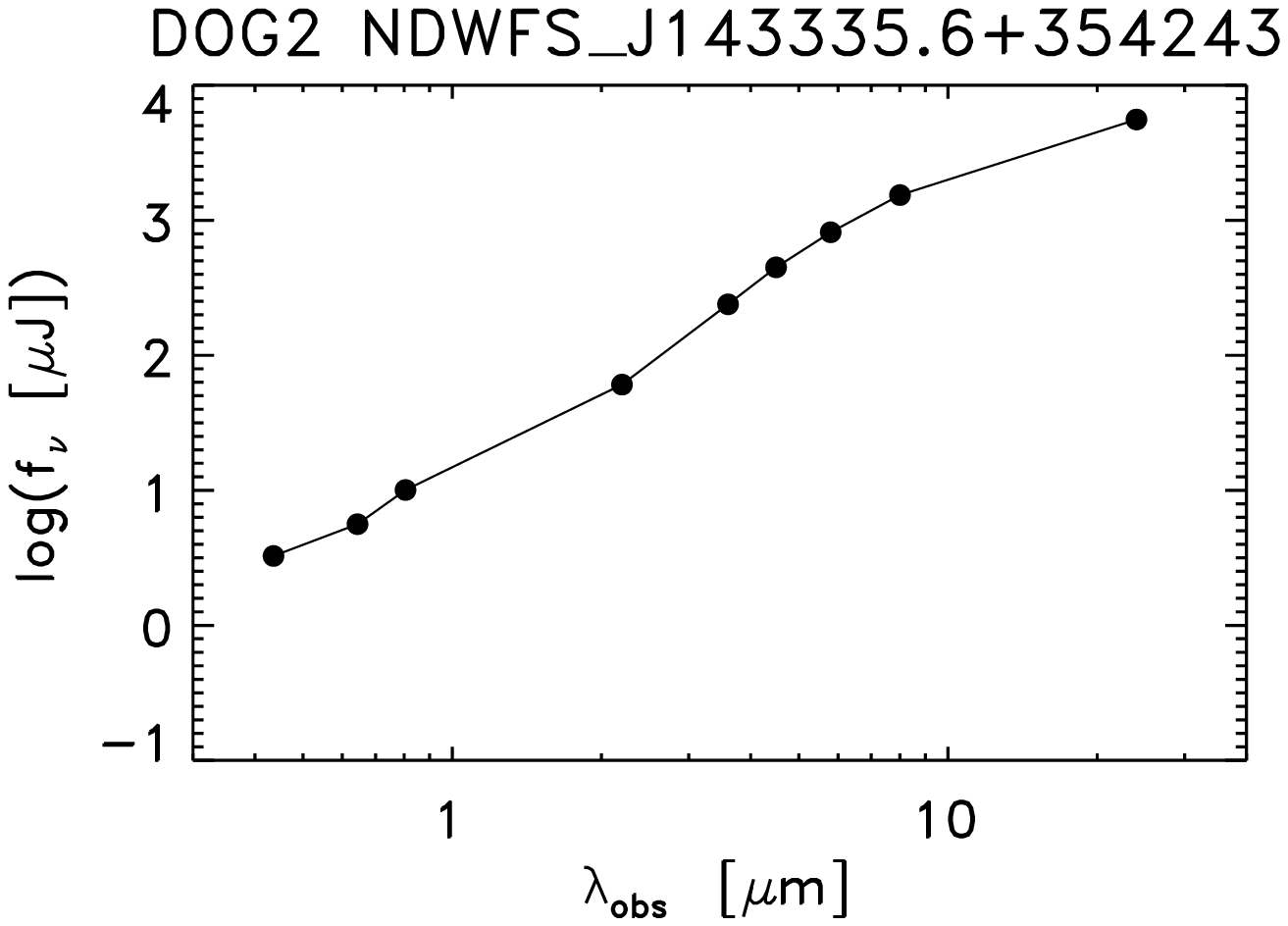}{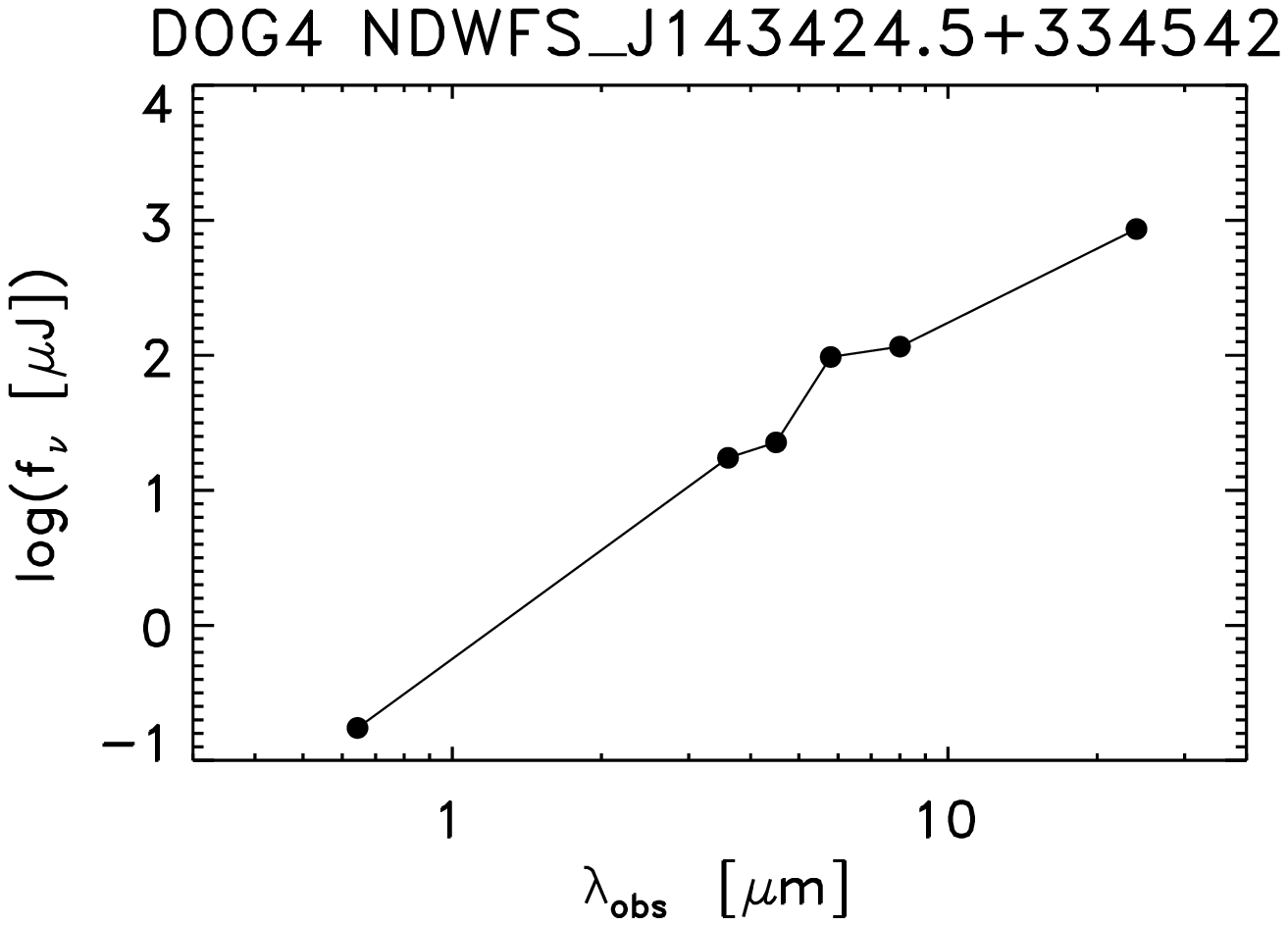}
\caption{\label{fig:SEDs} Optical through MIR SEDs of the four DOGs observed with OSIRIS.  These SEDs rise steadily into the MIR, and do not contain features such as a 1.6 \um\ stellar bump, or a Balmer break.  High redshift ULIRGs with SEDs like these have been generally shown to contain dust obscured AGN.}
\end{figure*}

The requirement of large \ha\ fluxes ($F_{H\alpha} > 1\times 10^{-16}$ ergs cm$^{-2}$ s$^{-1}$) was made to ensure that the OSIRIS observations were successful.  This flux limit translates to an $H\alpha$ luminosity of $5\times 10^{41}$ ergs s$^{-1}$ at $z=1$ and $3\times 10^{42}$ ergs s$^{-1}$ at $z=2$. This flux requirement also introduces a significant selection bias.  First, DOGs at redshifts where \ha\ falls between the $H$ and $K$-bands (i.e. $z\sim1.9$) will not be in the sample.  Many of the Bump DOGs lie in this redshift range because it places the 8 \um\ PAH feature into the Spitzer 24 \um\ band, reddening the the optical-to-MIR color.  Thus all of the DOGs in our sample have power-law SEDs (Figure \ref{fig:SEDs}) and are likely to be powered by AGN.  In addition, by selecting sources with the highest \ha\ fluxes, the sample skews to somewhat lower redshift than the typical DOG (which has $<z> \sim=1.9 \pm0.5$).  The final sample also skews to higher \tf\ flux with 3 of the 4 DOGs having $F_{\nu}(24) > 1$ mJy.     



 \begin{deluxetable*}{cccccccc}
\tabletypesize{\scriptsize}
\tablecaption{Keck Observation Run Summary \label{tab:obs}}
\tablehead{\colhead{Object Name} &\colhead{UT Date}   & \colhead{Exptime}  & \multicolumn{2}{c}{Filter}  & \multicolumn{2}{c}{Guide Star}& \colhead{Observing} \\
&&\colhead{[s]}&\colhead{Center [\um]} &\colhead{Width [\um]} & \colhead{$R$ [mag]} &\colhead{offset [$\arcsec$]} &\colhead{Conditions} }
\startdata
DOG1 & Jun 04, 2008 &  4500 	  & 1.571 & 0.078 & 16.3 & 34 &some cirrus \\
DOG2 & May 07, 2009 &  2700 	  & 1.504 & 0.075 & 15.1& 47 &clear \\
``                    & May 26, 2009 &  3600 	  & & & &&clear \\
``                    & May 27, 2009 &  3600 	  & & & &&clear \\
DOG3 & May 07, 2009 &  7200 	  & 1.765 &0.087 & 16.0 & 39& clear \\
DOG4 & May 27, 2009 &  10800	  & 2.175& 0.108 & 13.4 & 32&clear \\

\enddata			 
\end{deluxetable*}

\subsection{The Observations}
OSIRIS is an AO-fed instrument.  The Keck LGSAO system uses a deformable mirror to correct for atmospheric distortions to the wavefront, and thus recovers the diffraction limited resolution of the telescope ($\sim0.05\arcsec$ at 2.2 \um).  The system tracks atmospheric distortions with observations of a reference star within $50\arcsec$ of the science target (for tip-tilt correction), and a sodium laser guide star propagated from the telescope to the location of the target (for higher order corrections).  Table \ref{tab:obs} summarizes the conditions during OSIRIS observations, including magnitudes and separations of the tip-tilt guide stars.  

Observations taken in 2008 were made under photometric conditions with the instrument at the optimum well controlled temperature.  Unfortunately, during 2009, OSIRIS began to heat up, 5 degrees (K) above the optimal operating temperature.  The higher temperature produced noisier data and resulted in several issues that made 2009 data more difficult to reduce: the dark current was about 40\% higher compared to 2008, and warmer optics made extracting high signal-to-noise spectra more difficult. The observing conditions were clear during 2009 observing runs.  

In all cases, we observed with the $0.05\arcsec$ plate scale using narrow spectral filters centered on H$\alpha$, thereby allowing the largest possible field of view for the chosen plate scale.  The typical field of view was $2\arcsec \times 3\arcsec$ or roughly $20 \times 30$ kpc at these redshifts.

Because of the small field of view of OSIRIS, we first centered the AO tip-tilt guide star onto the OSIRIS frame.  We then manually offset to the target using offsets derived from NDWFS imaging.  This method provided good acquisition of the galaxy on the OSIRIS field of view.

Individual exposure times were 15 minutes with total on-source exposures ranging from 1 to 3 hours.  Small dithers were applied after each exposure.  However, the galaxy was always within the field of view of the instrument, and the galaxies were sufficiently small that no additional sky frames were needed.  

\subsection{Data Reduction}
The data were reduced with the OSIRIS reduction pipeline.  For each set of observations, a sky frame was created by median combining dithered science frames.  Each frame was sky subtracted, and cleaned of cosmic rays.  A 3D data cube was then extracted using rectification matrices supplied by Keck Observatory.  Frames were mosaicked together to produce a final 3-D data cube. 

As a result of running at higher temperature in 2009, optical elements within OSIRIS expanded, slightly altering the optical path.  Therefore the canonical rectification matrices used to convert the 2-D raw frames of overlapping spectra into 3-D data cubes were no longer valid.  While new rectification matrices were produced they did not work as cleanly as the original, and when applied to the data they created some pixels of unrealistically high and low count levels (e.g. $\pm10^{11}$ counts).  In order to remove these glitches we used a median combine (as opposed to a mean) when making the final data cubes.  
 
After each science target, an A0V star was observed with an identical instrument setup.  These observations were used to correct the spectra for Telluric absorption and to flux calibrate the images.  After the basic reduction of the standard star observations, a 1D spectrum was extracted of the star.   Hydrogen absorption lines were modeled out using the OSIRIS pipeline \emph{Telluric Extract} routine.  The stellar spectra were divided by a black-body spectrum with a temperature of an A0V star.  The resulting spectrum was then divided into the science data to correct for Telluric absorption.  

The spectra of the standard stars were also used to apply a rough flux calibration to the data.  A total flux for each star was measured from the OSIRIS data cube and compared to the expected flux of the star across the OSIRIS filter given the J, H, and K band fluxes \citep{Elias82}.  To determine the expected flux of the standard star in the OSIRIS filter we assumed a black-body spectrum across the filter band-pass, and assumed a flat filter function.  To check this calibration, we compare the flux in \ha\ for  DOG1 (See Table 1) to that found in \citet{Brand07} from long slit spectroscopy.  Our \ha\ measurement is 60\% larger than the Brand et al. value, which is not surprising as they suggest that there data could suffer slit losses on that order from pointing issues.  Because OSIRIS is effectively imaging in \ha, slit losses are minimal.

\begin{figure*}
\centering

\includegraphics[scale=0.7]{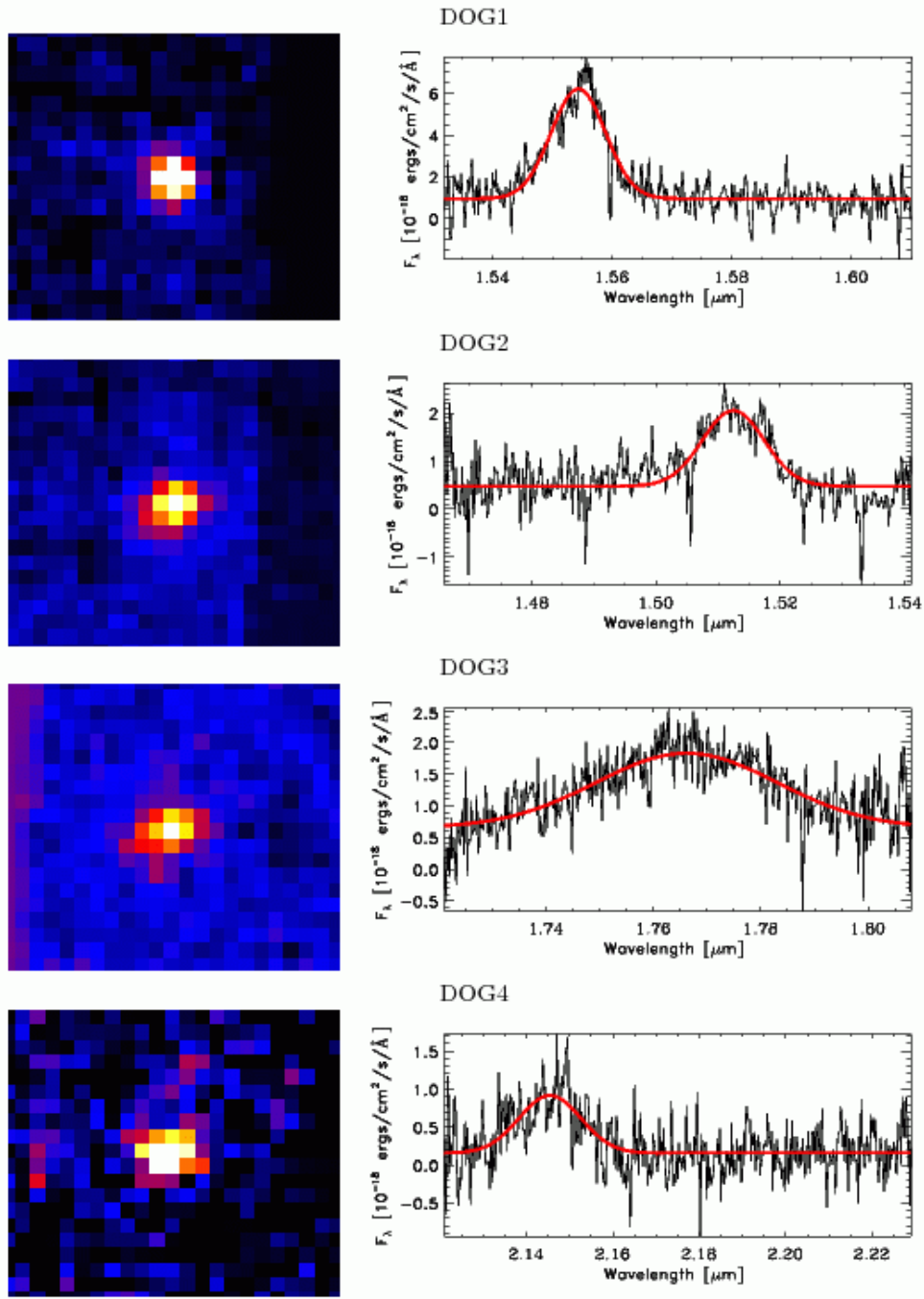}

\caption{\label{fig:Ha} \ha\ maps and 1D spectra of the sample ULIRGs.  Images are 1 arcsec on side ($\sim 10$ kpc at these redshifts).  All four DOGs show a point-like morphology in the \ha\ flux distribution.  This point-like flux is dominated by a broad \ha\ line, with FWHMs $> 2000$ km s$^{-1}$, suggesting AGN activity within the DOGs.  }
\end{figure*}


\begin{figure*}
\centering
\includegraphics[scale=0.7]{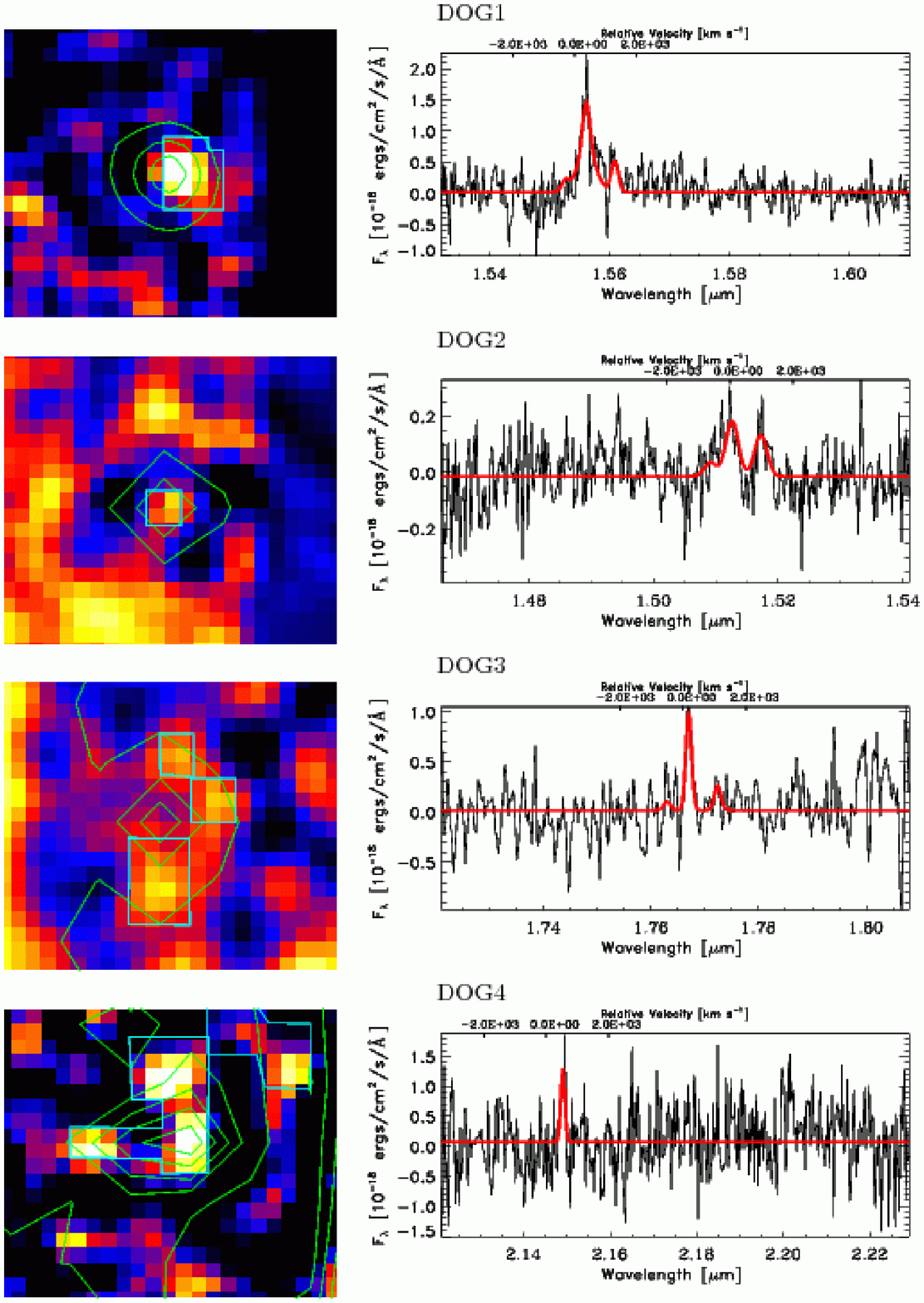}

\caption{\label{fig:narrow} Same as Figure \ref{fig:Ha} only now the broad line
component has been subtracted from the image, revealing the narrow-line \ha\
flux.  These images have been smoothed by a Gaussian kernel of 2 pixels width
to bring out the contrast.  Each DOG appears to contain one or more narrow-line
regions often spatially offset from the broad-line region (contours).  Cyan
boxes show the regions included in the extraction of the narrow-line flux.
These regions tend to be clumpy rather than extended (as might be expected in a
relaxed disk).  The top axis shows the velocity offset from the broad-line AGN
component. The spectra also show evidence for narrow [NII] lines.  }

\end{figure*}

\section{Spectral Line Measurements}

Figure \ref{fig:Ha} shows the \ha\ maps of the four DOGs observed by our program.  All show point-like \ha\ morphologies. We center a circular extraction region on the peak of the \ha\ flux, and extract a region with radius of $0.15\arcsec$, enclosing the diffraction limited core of the PSF which is measured to have a FWHM of $0.1\arcsec$.    The spectra of these point-like objects  (Figure \ref{fig:Ha}) show broad \ha\ emission.  We first fit these broad lines with a single Gaussian profile (red curve in Figure \ref{fig:Ha}).  The resulting fits have line widths in excess of 2000 km s$^{-1}$  indicative of AGN activity.  While the spectra are noisy, we also attempt a fit that includes narrow-line \ha\ and [NII $\lambda$6548,6583].  The total fluxes and line widths of the broad \ha\ line do not change significantly with the inclusion of these narrow-lines in the fit.  Fluxes and line-widths of the broad \ha\ emission are given in Table \ref{tab:bha}.


We also attempt to identify the spatial extent of the narrow-line emission in each system.   To accomplish this, we first subtract off the broad-line AGN component.  We create a point-spread-function (PSF) image by collapsing the data cube in the spectral region that encompasses the wings of the broad-line \ha\ spectrum.  At each wavelength, we scale the peak of the PSF image to the flux of the modeled broad \ha\ line and subtract it from the data cube.  The result of this process is a 3-D data cube of the narrow \ha\ flux with the broad-line \ha\ removed.    

The narrow \ha\ images of each galaxy are shown in Figure \ref{fig:narrow}.  These images have been smoothed by a Gaussian kernel with FWHM of 2 pixels to enhance the detection of any prominent sources.  In general, these images do not show regular morphologies such as those expected if star formation were smoothly distributed within well-behaved disk structures.    Rather, these regions appear as individual clumps.  Except for  DOG2, these narrow-line regions are separated from the nuclear region which hosts the AGN  (indicated by green contour lines).    

Shown on the images are the regions used to measure the narrow \ha\ line widths and strengths.  Some regions that appear bright in the narrow \ha\ images were not included in the  creation of the final spectrum.  Individually, the spectra of these regions showed no indication of containing emission line flux.  Rather, these regions appear to be noise fluctuations in the collapsed 2D images.

The spectra of the narrow-line regions are shown in Figure \ref{fig:narrow}.  Evidence for narrow \ha\ emission is found in each DOG.  We fit the spectra for \ha\ and [NII] lines, using Gaussian profiles (red line in Figure).  FWHMs of the narrow emission lines vary from $\sim200-500$ km s$^{-1}$, with typical fluxes from $(0.5-2.5)\times10^{-17}$ ergs cm$^{-2}$ s$^{-1}$.  The velocity offsets between the narrow and broad \ha\ lines in each galaxy range from $-150$ to 500 km s$^{-1}$.

Three of the DOGs show evidence for [NII] emission.   DOG1 and  DOG3 have [NII]/[\ha] line ratios of $\sim0.3$.  In contrast,  DOG2 has [NII]/[\ha]$=0.74$.  Such a high ratio likely indicates that the production of [NII] in this galaxy is being enhanced by AGN activity \citep{Swinbank04}, a reasonable assumption given that the emission is located spatially atop the nuclear broad \ha\ flux. Shocks in the nuclear gas could also result in high [NII]/[\ha] ratios.  The narrow-line measurements are summarized in Table \ref{tab:nha}. 
 
\section{Properties of the Sample}

\subsection{Dust Extinction}
In nebular theory, the \hahb\ ratio is fixed by the radiation field. For case B recombination, in a density-bounded HII region powered by star formation,  \hahb$=2.86$.  For AGN-ionized regions, \hahb$=3.1$ \citep{Osterbrock89}.  \hahb\ ratios larger than these canonical values indicate the presence of dust, which will preferentially attenuate  the bluer \hb\ line compared with the \ha\ line.  Thus, a measurement of the \hahb\ line ratio can be used to estimate the reddening in a system.  

The OSIRIS observations of the DOGs in our sample were only made in a small spectral window ($\sim100$ nm wide) about \ha\ and therefore do not contain the information necessary to produce an estimate of dust extinction.  However, 3 DOGs in the sample were observed previously with NIR long-slit spectrographs covering both \ha\ and \hb\ lines.  While \hb\ was not strongly detected in any of these observations, limits on the line strengths were obtained.  

DOG1 was observed with the Gemini NIRI spectrograph and the results were published in \citet[][object J143027.1+344007]{Brand07}. In this  system, the broad-line Balmer decrement \hahb\ $\geq 22.5$, which, assuming a Milky Way dust curve \citep{Cardelli89}, translates to a reddening, $E(B-V) \geq 2.09$ mag, or $A(H\alpha) \geq 4.86$ mag.  Thus the extinction corrected \ha\ emission in this system is likely to be a factor of 70 larger (or more) than measured.

 DOG3 and  DOG2 were observed by Palomar TripleSpec (Melbourne et al. in preparation).  Unfortunately, the limits on the \hahb\ ratios in the Palomar data are less stringent than the NIRI data, with \hahb\ $\geq 6.5$ and 7.6 respectively.  These are equivalent to $A(H\alpha)\geq 1.84$ mag and 2.22 mag, or roughly a factor of 5 - 10 in flux. However,  $A(H\alpha)$ could be significantly larger then this lower limit. 

For  DOG4 we do not have a measurement of H$\beta$, and therefore do not have a measure of the dust extinction.  We choose to adopt the minimum dust extinction measured from the other 3 DOGs ($A(H\alpha) \geq 1.84$ mag) as a lower limit on the extinction for  DOG4.  


\subsection{Black Hole Masses}
Under certain assumptions the OSIRIS data cubes can be used to estimate the black hole masses in our sample galaxies.  To make the estimate we require two measurements: (1) the \ha\ line width, and (2) a continuum flux from the AGN point source.   The first quantity we measure directly from the spectra.  Unfortunately, the second quantity is difficult to measure directly from the OSIRIS data.  While continuum is seen in the spectrum of the central region of the galaxy, it is not clear how much of this continuum is produced by the AGN and how much is produced by the galaxy.  However, \citet{Greene05} show that there is an empirical relationship between broad-line \ha\ flux and the continuum flux of the AGN,
\begin{equation}
L5100=1.23\times 10^{7}\cdot (L_{H\alpha})^{0.864}
\end{equation}
We use the reddening-corrected broad-line \ha\ luminosities to determine the AGN continuum luminosity at 5100 \AA, L5100. 
With the \ha\ line width and continuum flux, we calculate the black hole mass from the empirical calibration of \citet{Peng06}, \vspace{1cm}
\begin{eqnarray}
 \label{eqn:bh}
\lefteqn{M_{BH}  =  9.7\times10^{6} \cdot}\nonumber\\
& &  \left(\frac{\textrm{L5100}}{1\times10^{44}
  \textrm{[ergs/s]}}\right)^{0.59}\cdot  \left(\frac{\textrm{line-width}}{1000
  \textrm{[km/s]}}\right)^{2.06} \; M_{\odot} 
\end{eqnarray}
The estimated black hole masses are summarized in Table \ref{tab:bha}; they range from $(1-9)\times 10^8$ M$_{\odot}$.  DOG3 hosts the most massive BH, not surprising given the width of the broad \ha\ line in that system. The intrinsic scatter in this empirically derived relation (Equation \ref{eqn:bh}) is about $0.3-0.4$ dex \citep[e.g.][]{Vestergaard02}.  However, systematic uncertainties in the dust correction could skew these results to lower BH masses and all reported BH masses are lower limits.

\begin{deluxetable*}{ccccccc}[ht]
\tabletypesize{\scriptsize}
\tablecaption{Broad-Line H$\alpha$ Properties \label{tab:bha}}
\tablehead{
\colhead{Object Name}  & \colhead{Obs. H$\alpha$ flux} & \colhead{H$\alpha$ width} & \colhead{H$\alpha$ Lum.\tablenotemark{a}} & \colhead{$A[H\alpha]$} & \colhead{Black Hole Mass\tablenotemark{b}} &\colhead{Host Galaxy}\\
& \colhead{[$10^{-16}$ ergs cm$^{-2}$ s$^{-1}$]} & \colhead{[km s$^{-1}$]} & \colhead{[$10^{42}$ ergs s$^{-1}$]} & \colhead{[mags]} & \colhead{[$10^6$ M$_\odot$]}&\colhead{M$_{R}$\tablenotemark{c}}}
\startdata
DOG1 & 5.98 $\pm0.25$ & $2068\pm22$ & 6.77 &$\ge$4.86 \tablenotemark{d} & $>490$ &-20.06\\
DOG2 & 1.97 $\pm0.09$ & $2288 \pm 40$& 1.98 &$\ge$2.22 \tablenotemark{e} & $>103$ &-23.85\\
DOG3 & 4.97 $\pm0.15$ & $6757\pm 96$& 2.60 &$\ge$1.84 \tablenotemark{e} & $>878$ &-24.73\\
DOG4 & 1.36 $\pm0.20$ & $2358 \pm221$& 5.34 &$\ge$1.84 \tablenotemark{f} & $>138$&-24.30 \\

\enddata			 
\tablenotetext{a}{Uncorrected for dust attenuation} 
\tablenotetext{b}{Corrected for dust attenuation}
\tablenotetext{c}{Corrected for AGN contamination, but no dust correction}
\tablenotetext{d}{\citet{Brand07}}
\tablenotetext{e}{Melbourne et al. in Preparation}
\tablenotetext{f}{No measurement of H$\beta$ available so used the minimum correction from the others in the sample.}
\end{deluxetable*}

\subsection{Rest-Frame Optical Galaxy Stellar Luminosity}  
To better compare the DOGs with other galaxy samples (Section 5) we attempt to measure the stellar contribution to the galaxy luminosity, independent from the AGN contribution.  Unfortunately, because the spatially extended galaxy continuum is not well detected in the OSIRIS frames, we can not use the OSIRIS data to constrain the relative contribution of AGN and star-light to the total luminosities.  However, we get some hints from high spatial resolution \HST\ and AO imaging of larger DOG samples, which show that the AGN (central point source) typically contributes only $\sim$10\% of the galaxy light at NIR (rest-frame optical) wavelengths \citep{Melbourne09, Bussmann09}.

While this ratio could prove to be a useful rule of thumb for DOGs, we know that it does not hold universally.  In fact, we have a high spatial resolution AO image of  DOG1, which is actually point-source dominated \citep{Melbourne09}, suggesting that the AGN could be contributing a much higher fraction of the light in this system.  This example suggests that in the absence of imaging we should attempt some other method for determining the AGN contribution to the galaxy luminosity.

Fortunately, the OSIRIS data can be used to measure the AGN continuum contribution based solely on the \ha\ flux.   As quoted from Greene \& Ho, the flux from \ha\ correlates with the AGN continuum emission at 5100\AA (L5100). With a measure of the AGN continuum in hand, we convert to total AGN flux in a given passband assuming a standard QSO spectrum.  Subtracting the AGN flux from the total observed flux (at rest-frame optical wavelengths), we derive the stellar luminosity of each galaxy which we convert into a rest-frame R-band absolute magnitude assuming a power-law spectral energy distribution for the galaxy light.  As expected from its high spatial resolution AO image, the flux of  DOG1 is dominated by AGN light; the AGN contributes 95\% of the total.  However, the other DOGs show significantly less contribution from AGN light to the total luminosity, $\sim10$\%, which is typical for DOGs based on results of image decomposition.   

Un-corrected for reddening, the DOGs in our sample have $M_R = -20$ to $-24$.  However, DOGs are among the most dust obscured galaxies known, with colors redder than the typical low-z ULIRGs.  This suggests that even the galaxy light may be heavily attenuated by dust.  If we apply the dust correction from the broad-line regions to the entire galaxy (likely an upper limit on the actual dust obscuration for the galaxy) then the DOGs are very luminous, $M_R = -24$ to $-26.5$.   


\subsection{Star Formation Rates}
After subtracting the broad-line \ha\ component, each DOG shows additional narrow-line \ha\ flux.  If we assign all of the observed narrow \ha\ flux to star formation we can estimate the star formation rate (SFR) of each galaxy.  We use the prescription in \citet{Wright10} adopted from \citet{Kennicutt98}, 
\begin{equation}
SFR\; \textrm{[M}_\odot/\textrm{yr}]=L(H\alpha)/2.5\times10^{42}
\end{equation}
Uncorrected for reddening, we derive typical star SFRs for the DOGs of only $0.5-2.0$ M$_\odot$ \yr (Table \ref{tab:nha}).  

\begin{deluxetable*}{ccccccccc}[ht]
\tabletypesize{\scriptsize}
\tablecaption{Narrow-Line \ha\ Properties \label{tab:nha}}
\tablehead{
\colhead{Object Name}  & \colhead{H$\alpha$ flux} & \colhead{H$\alpha$ width} &  \colhead{\ha\ Lum \tablenotemark{b}} & \colhead{SFR\tablenotemark{b}} & \colhead{[NII]/\ha} & \colhead{Metallicity\tablenotemark{c}} \\
& \colhead{[$10^{-17}$ ergs cm$^{-2}$ s$^{-1}$]} & \colhead{[km s$^{-1}$]}  &\colhead{[$10^{41}$ ergs s$^{-1}$]} & \colhead{[M$_\odot$ yr$^{-1}$]}& & [12+log(O/H)]}
\startdata
DOG1 & $2.5\pm0.2$& $533 \pm 47$ & 2.80& $1.3 \pm0.1$ &$0.27\pm0.03$&8.6\\
DOG2\tablenotemark{a} & $0.50 \pm0.12$ & $470 \pm 48 $& 0.50& $0.22 \pm0.02\tablenotemark{a}$ &$0.74\pm 0.24$ &9.1\tablenotemark{a}\\
DOG3 & $1.3 \pm0.1$& $210 \pm 56$ & 2.41& $1.1\pm0.1$ &$0.29 \pm 0.04$&8.7\\
DOG4 & $1.4\pm0.1$& $152 \pm80$& 5.53& $2.5 \pm0.1$& - &\\

\enddata			 
\tablenotetext{a}{The narrow lines observed for TDOG\_5 arise from the same region as the broad lines, and the line ratios indicate that these lines are produced by the AGN rather than star formation.  As a result the calculated star formation rates and metallicities for TDOG\_5 are likely incorrect.}
\tablenotetext{b}{Uncorrected for dust attenuation.}
\tablenotetext{c}{Derived from the [NII]/\ha ratio \citep{Melbourne02}, with an uncertainty of roughly 0.3 dex.}
\end{deluxetable*}

As with the broad-line fluxes, these narrow fluxes are likely to be heavily dust extincted.  Because they are spatially removed from the center of each galaxy, the dust obscuration may be different than for the AGN. However, we can use the AGN obscuration as a rough estimate.  Doing so gives SFRs 1-2 orders of magnitude larger than measured.  Even with these correction factors the star formation rates that are measured are small for ULIRGs, $< 100$ M$_{\odot}$ \yr.  In addition, some of the narrow-line flux could be produced by the AGN, meaning that the SFR could be significantly lower than these limits.

In addition, we have shown that we are sensitive to very small star formation rates.  Clearly the surface density of star formation in any extended component must be small, e.g. $<0.5$ \sm \yr  kpc$^{-2}$ (before correcting for dust attenuation), or our observations would have detected it. 

\subsection{Metallicity}
Three galaxies in our sample show evidence for an [NII $\lambda$6583] collisionally excited emission line.  As shown in \citet{Melbourne02}, the [NII]/\ha\ line ratio can be used as a rough proxy for metallicity in star forming galaxies.   Assuming the [NII] emission in these systems is entirely the product of star formation we can estimate the metallicity of these systems using,
\begin{eqnarray}
12+\textrm{log(O/H)} & = & 9.26+1.23\cdot\textrm{log([NII]/H}\alpha)+\nonumber \\
& & 0.204\cdot[\textrm{log([NII]/H}\alpha)]^2
\end{eqnarray}

For the three galaxies for which [NII] is observed (DOGS 1, 2, and 3), the estimated metallicity
is roughly solar or larger.   However, for  DOG2, the [NII] emission lies atop
the nucleus of the galaxy (Figure 2) and is likely boosted by AGN heating.  
In this system, [NII]/\ha$>0.7$, another indication that the
[NII] line may be enhanced by AGN activity \citep{Brand07}.  Therefore, 
Equation 4 is expected to over-predict the
metallicity in DOG2.    The metallicities are given in Table 4. 

\subsection{Kinematics}

Ideally, we would have detected \ha\ across an extended star forming disk in each galaxy.  We could then have then used the observed kinematics to model the mass distributions within the DOGs.  However, due to poor S/N and/or the nature of the DOG kinematics, none showed evidence for well ordered rotation in an extended gas disk.  

Despite these issues, there do appear to be narrow \ha\ emitting regions detected in each galaxy.  In DOGs 1 and 2, the narrow \ha\ flux arrises from a single source near the nucleus, each only $1-2$ kpc in diameter.  The line-widths of the narrow \ha\ emission in these two DOGs are resolved at 530 and 470 km s$^{-1}$ respectively.  If the ionized gas is tracing a virialized mass distribution, the masses within these knots are extremely large, M $>10^{11}$ \sm.  More likely, the gas is not virialized. The line-widths could be enhanced by the nearby AGN, or from kinematically disturbed gas, or from large bulk motions such as from expanding bubbles from a wind. 
In the case of  DOG2, the narrow emission resides exactly atop the location of the broad-line region, suggesting some fraction of the narrow-line flux may be from the AGN.  The high [NII]/\ha\ line ratio of this region also suggests AGN contamination.  For  DOG1, the [NII]/\ha\ line ratios are more consistent with star formation.  

The other two galaxies,  DOG3 and  DOG4, appear to show extended narrow \ha\ emission in several distinct knots with separations as far as 5 kpc.  Each of the knots has a very small \ha\ line width of $ \sim150-200$ km/s, and there are no large velocity gradients across these multiple emitting regions.

In each case, the central wavelength of the narrow-line component is offset from the broad-line component.  These kinematics offsets range from 150 km s$^{-1}$, in DOGs 2 and 3, to 350 and 500 km s$^{-1}$, in DOGs 1 and 4 respectively.

\begin{figure*}
\centering
\includegraphics[scale=0.5]{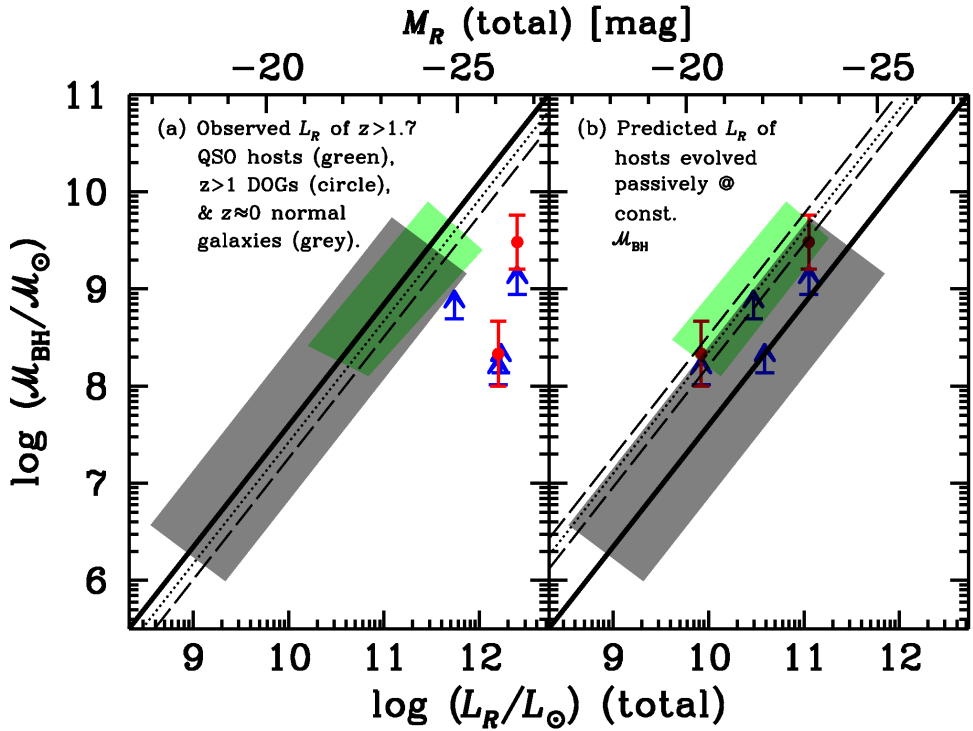}

\caption{\label{fig:bh} Left: Estimated black hole mass vs. host luminosity for
the DOGS (blue limits), $z\sim2$ AGN \citep[green
shading][]{Peng06,Ridgway01,Kukula01} and local galaxies (grey shading).  Also shown are X-ray derived BH masses for two of the DOGs (red points).  In both cases the X-ray results return BH masses that are larger than the OSIRIS derived lower limits. The
best fit local relation is shown as a solid black line, while the best fit $z\sim2$
relation is shown as a dotted line; the dashed lines show different SED
assumptions used to derive the host galaxy luminosities.  Both the BH mass
estimates and the galaxy luminosities of the DOGs have been corrected for dust
reddening.  No reddening correction is given to the other systems. However,
those corrections are expected to be small.  Given the high intrinsic
luminosities of the DOG host galaxies, their black holes appear under-massive
compared with the $z\sim2$ and local samples.  These differences can be explained
if the DOGs contain significant young stellar populations which boost the host
galaxy luminosity.    Right: the luminosity expected for the DOGs and other $z\sim2$
galaxies after accounting for the fading of their stellar populations.  An
instantaneous burst fading, with $z_{\rm formation}$ at the observed redshift,
was applied to the DOGs while passive fading was applied to the $z\sim2$ galaxies.
While the DOGs fade to the local relation, the $z\sim2$ systems fade past the
local relation, suggesting that they need to undergo additional stellar mass
growth to eventually land on the local relation \citep{Peng06}. Assuming an
Eddington accretion of $\sim2$ \sm yr$^{-1}$, the DOGs could easily grow their
BHs onto the $z\sim2$ relation (after accounting for fading).  } 

\end{figure*}

\begin{figure*}
\centering
\includegraphics[scale=0.5]{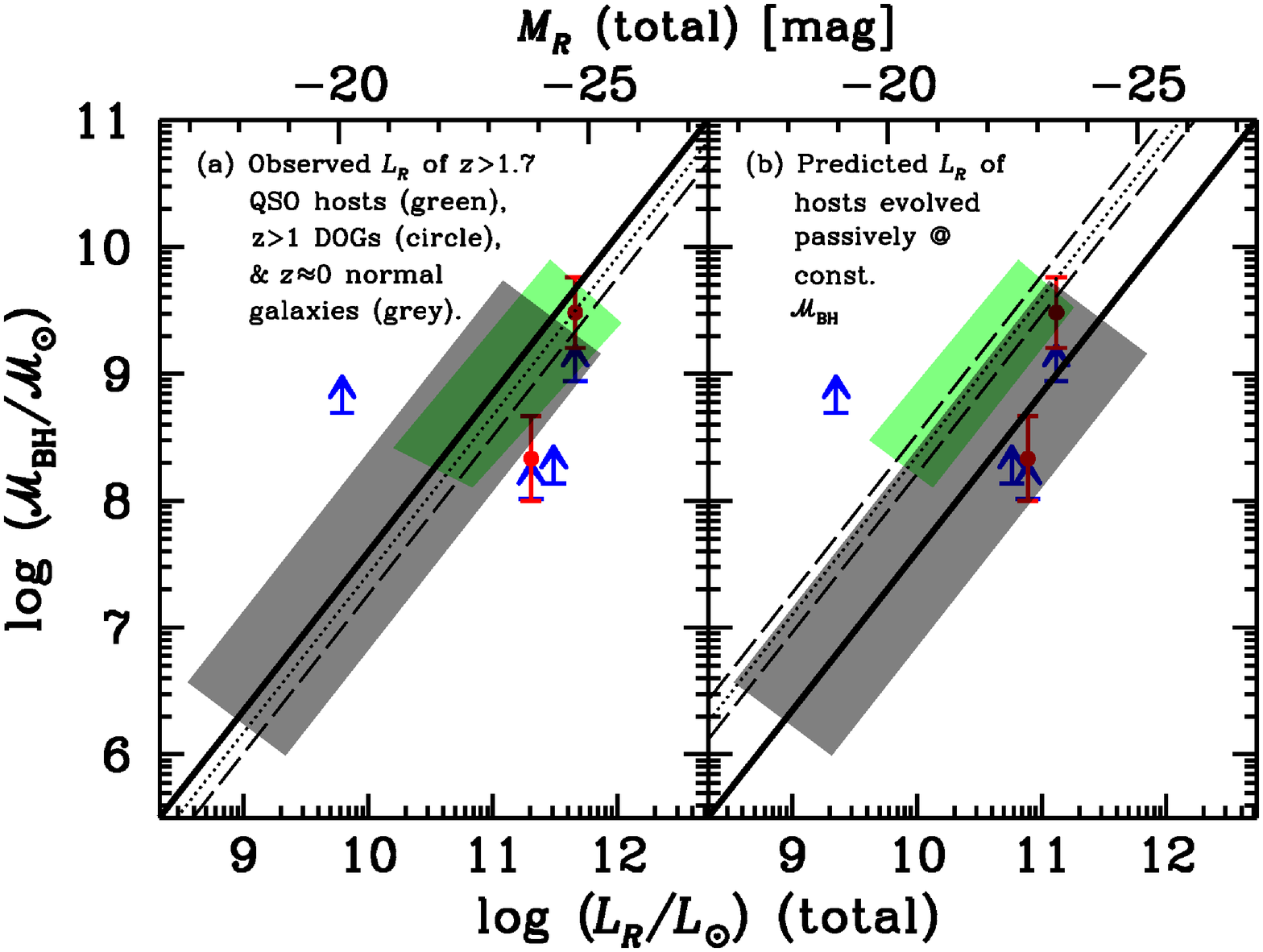}

\caption{\label{fig:bh2} Left: same as Figure \ref{fig:bh} only now no dust
correction has been applied to the DOG host galaxy luminosities.  Right: now
shows the DOGs fading passively with an assumed epoch of star formation at redshift, $z=5$ --- the same prescription applied to the other $z\sim2$
AGN hosts.  In this scenario, the bulk of the DOGs again fade to the local
relation. } 

\end{figure*}

\section{Discussion}




In this set of four DOGs, the broad \ha\ lines associated with AGN have fluxes
10-20 times larger than the detected narrow lines produced in star forming
regions.  The relative strengths of
these lines suggest that the bulk of the dust heating results from AGN activity
rather than star formation.  Based on the widths of the broad lines, and the
continuum flux of the nuclear region, we estimate lower limits on the black hole masses for the
four DOGs in our sample, in the range of $(1 - 9) \times10^8$ \sm.

\subsection{Comparing the BH Masses of DOGs to BH Masses of Other $z\sim2$ Galaxies}

Figure \ref{fig:bh} compares the BH masses of the the DOGs (blue limits) with
the BH masses in other $z\sim2$ and $z\approx0$ galaxies
\citep{Peng06a,Ridgway01,Kukula01}.  The $z\sim2$ galaxies are a mix of radio-loud
and radio-quiet quasars.  Their black holes masses were measured with a
technique similar to the one used in this paper, except that the emission lines
used were from the rest-frame UV, C~IV $\lambda 1549$ and  Mg~II $\lambda 2798$ 
\citep[][]{Peng06}.  The $z\sim2$ host galaxy magnitudes of the
quasars were measured from high spatial resolution NIR imaging from HST.  The
comparison sample of interest at $z=0$ are elliptical galaxies with BH masses
measured from high spatial resolution HST spectroscopy of the circum-nuclear
gas and stars in each galaxy \citep[e.g][]{Kormendy01, Ferrarese05}, where the
R-band quantities are taken from \citet{Bettoni03}.  

When placing the DOGs into Figure \ref{fig:bh}, we must choose what level of dust correction to apply to the BHs and host galaxies.  For the BHs themselves we show the dust corrected lower limits on the masses.  For the host galaxies, we use two different prescriptions.   In  Figure \ref{fig:bh} we adopt the dust corrections derived from the nuclear regions.  While these are lower limits on the dust extinction of the BHs they could be reasonable upper limits on the galaxy extinction.  Then in Figure \ref{fig:bh2} we take the other extreme, no dust correction on the galaxy host luminosities.  We discuss both hypotheses below.

Figure \ref{fig:bh}  shows that the BH masses measured for the DOGs are similar to the $z\sim2$ AGNs. However, as a function of galaxy luminosity, the DOGs appear in a different region of the BH-mass/galaxy-luminosity plot compared with the $z\sim2$ and local
samples.  As a function of black hole mass, their extinction corrected host
luminosities are much higher (Figure \ref{fig:bh}a).  This means that either
(1) the DOGs have under-massive black holes, (2) their host galaxy luminosities are
enhanced compared with the other galaxy samples, or (3) the dust corrections are incorrect.  If we take the dust corrections as measured, and make the
assumption that the DOGs must land on the local relation today, they can reach
this relation by a combination of fading, and/or black hole growth. A similar conclusion was reached to explain a sample of $z\sim2$ SMGs, also with smaller than expected black holes for their host stellar masses \citep[estimated from rest-frame NIR fluxes,][]{Borys05}

The simplest way to explain these differences is if the DOGs have larger
fractions of very young stellar populations that fade rapidly with time.  The
DOGs may have had significant recent star formation, contributing to their high
luminosities and large dust reservoirs.  In contrast, the local galaxies
generally have old stellar populations that have already faded from their
high-$z$ luminosities.  If we assume that the DOGs fade as an instantaneous
burst from the time of their observation at $z\sim1.5$ \citep[e.g.][]{van-Dokkum01}, 
then in fact they will fade onto the local BH mass galaxy luminosity
relation, as shown in Figure \ref{fig:bh}b.  This fast fading scenario is
appropriate if most of the mass is formed at $z\sim1.5$.

\citet{Peng06} provide another possible evolutionary scenario for the DOGs on the BH-mass/galaxy-luminosity plot.  Following the prescription that they used for the $z\sim2$ AGN in Figures \ref{fig:bh} and \ref{fig:bh2}, we can assume instantaneous star formation
at $z=5$, followed by passive evolution.  The essence of this model lies in the
observation that dusty galaxies with young stellar populations generally are
more luminous than dust-free galaxies with old populations of equal stellar
mass.  Therefore, passively fading the {\it observed} star light at $z\approx2$,
without correcting for extinction, should give an upper limit on the light
produced by the dominant stellar mass component by $z=0$.  We show the result
of this model in Figure \ref{fig:bh2}, where the galaxy stellar
luminosities are significantly fainter than Figure \ref{fig:bh}, because we
do not correct for dust extinction in the galaxy.  Even under these assumptions, 
three of the four DOGs
still have larger galaxy luminosities at a given BH mass compared with the
local and \citet{Peng06} high-$z$ samples (Figure
\ref{fig:bh2}a).  If we then apply a passive fading rate, as was applied to the
other $z\sim2$ AGN, the bulk of the DOGs again fade to the local relation (Figure
\ref{fig:bh2}b).

In either dust correction scenario, the bulk of the DOGs fade onto the local
BH-mass/galaxy-luminosity relation by today.  This behavior is different from
the $z\sim2$ quasar host galaxies which actually fade past the local relation,
even when the fading applied is only passive fading.  If the $z\sim2$ AGN hosts
faded as instantaneous bursts (as has been done with the DOGs in Figure 3) they
would be even further off of the local relation. This phenomenon is discussed
extensively in \citet{Peng06}; the $z\sim2$ AGN host galaxies need to grow in
stellar mass to land on the BH-mass/galaxy-luminosity relation today.  Peng et al. estimate 
that a specific star formation rate of 1.2 Gyr$^{-1}$ would place the quasar hosts onto the local relation by $z=1$ or, for longer star formation time scales,
the specific star formation could be as low as 0.6 Gyr$^{-1}$. 

One possible explanation for why the DOGs behave differently from the
high-$z$ comparison samples is that we have under-estimated 
the BH masses by under-estimating the dust corrections on \ha. 
As a check on the BH masses measured from the OSRIS data, we use the 5 Ks Chandra X-ray observations of the \boot\ field to make another BH mass estimate. These observations are relatively shallow but two of DOGs in our sample are detected with $> 10$ counts.  For these galaxies, we derived the $2-10$ keV luminosities using an absorbed power-law model as the proxy. We fixed the spectral index, Gamma$=2$, and derived the column density from the hardness ratios, taking Galactic columns into account.  A summary of the fitted parameters is given in Table \ref{tab:xray}.  We then use the empirical relationship between AGN X-ray luminosity and optical continuum flux given by \citep{Maiolino07},

\begin{equation}
\textrm{log}(L5100)=\frac{\textrm{log}(L_{2-10\textrm{keV}}) -11.78}{0.721},
\end{equation}
to derive a second estimate of $L5100$. Using the X-ray derived $L5100$, we recalculate the BH masses from Equation \ref{eqn:bh}. The X-ray derived BH masses are given in Table \ref{tab:xray}.  In both cases the BH masses derived from the X-ray observations are larger than the lower limits derived from the optical OSIRIS data (with large uncertainties).  The X-ray derived BH mass for DOG2  is estimated to be $1.5 - 2.5$ times the lower limit from the OSIRIS data, while DOG3 has a BH mass about $2-3$ times larger. 

\begin{deluxetable*}{ccccccc}
\tabletypesize{\scriptsize}
\tablecaption{X-ray Derived Properties \label{tab:xray}}
\tablehead{
\colhead{Object Name}  & \colhead{X-ray} & \colhead{0.2-10 keV Flux\tablenotemark{a}}  & \colhead{0.2-10 keV  Lum.\tablenotemark{b}} & \colhead{HI Column} & \colhead{2-10 keV X-ray Lum.\tablenotemark{c}} & \colhead{Est. BH Mass}\\
& \colhead{Counts} & \colhead{[$10^{-14}$ ergs cm$^{-2}$ s$^{-1}$]} & \colhead{[$10^{44}$ ergs s$^{-1}$]} & \colhead{$10^{21}$ cm$^{-2}$} & \colhead{[$10^{44}$ ergs s$^{-1}$]} & \colhead{[$10^6$ M$_\odot$]}}
\startdata
DOG2 &  $16 \pm 5$& $4.4 \pm1.5$ &  $4.2 \pm 1.5$ & 6.1 & $2.1 \pm 0.7$ & $215 \pm 72$ \\
DOG3 &  $18 \pm 6$& $4.9 \pm1.7$ &  $8.8 \pm 2.9$ & 0.6 & $3.9 \pm 1.3$ & $3040 \pm 850$
\enddata
\tablenotetext{a}{Observed-frame}			 
\tablenotetext{b}{Rest-frame, uncorrected for internal absorption} 
\tablenotetext{c}{Rest-frame, corrected for internal absorption}
\end{deluxetable*}

While the X-ray results push DOGs 2 and 3 to higher black hole masses, it is not clear how the estimated host galaxy luminosities should change, if at all.  The X-ray results suggest that  $A_V$ has been underestimated for these two DOGs; not a surprise given the week constraint from the TripleSpec spectra on the H$\alpha$/H$\beta$ ratios.  If we assume that the additional reddening is primarily affecting the central BH, then the AGN (even though they are intrinsically brighter than the OSIRIS-based estimate) should continue to contribute only a small fraction of the rest-frame optical light of the galaxies, because of the higher extinction. Thus the host galaxy luminosities could remain unchanged.  After accounting for luminosity evolution within the host galaxies (see Figures \ref{fig:bh}b and \ref{fig:bh2}b), the X-ray derived BH masses place the DOGs closer to their $z\sim2$ counterparts in the BH mass/galaxy luminosity plane, although they are also still consistent with fading onto the local relation.

In all of these scenarios the black holes within the DOGs may continue to grow. With roughly Eddington accretion rates of 2 M$_{\odot}$/yr, the DOGs could easily reach the $z\sim2$ relation in $50 - 100$ Myr.   Of course if their BHs were to grow onto the $z\sim2$ relation, the DOGs would then also need to continue to grow their stellar mass to reach the local relation by today.

\begin{figure}
\includegraphics[scale=0.6]{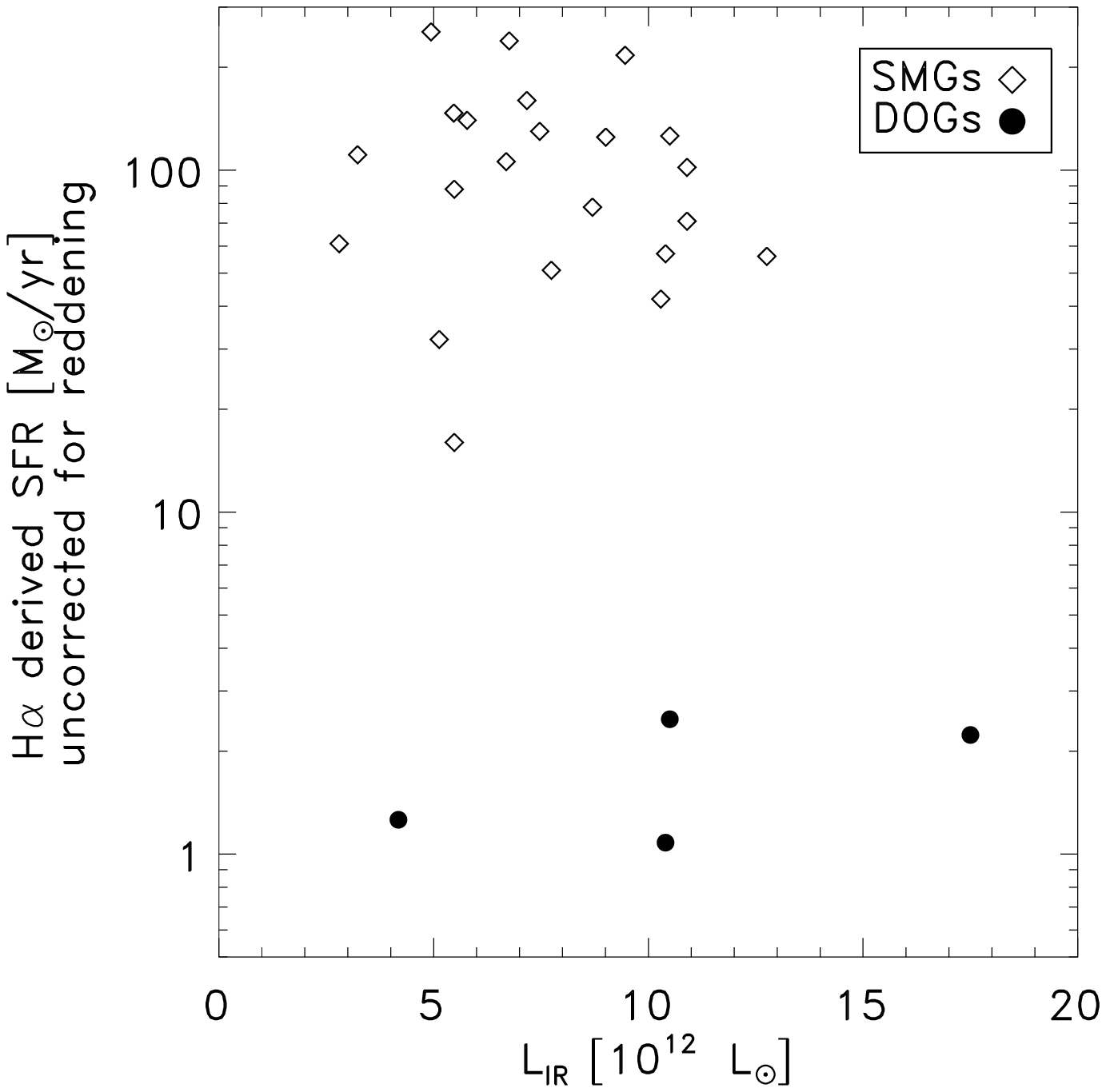}

\caption{\label{fig:SFR} The \ha\ derived SFRs of the DOGs and a comparison
sample of $z\sim2$ SMGs from \citet{Swinbank04}, plotted against $L_{IR}$.  No
dust attenuation correction has been applied to the SFRs of either sample. 
Clearly, these four DOGs show significantly less SF than the
SMGs despite having similar IR luminosities. Note: $L_{IR}$ estimates of
the DOGs are based on \tf\ fluxes and \citet{Chary01} models.  They are only
meant to be representative and do not reflect actual measurements of the FIR
fluxes of the DOGs.}

\end{figure}

\subsection{Star Formation Rates, Metallicities and Kinematics}

The apparent (i.e., uncorrected for reddening) star formation rates of the DOGs in our sample
are surprisingly small, $<2$ M$_{\odot}$ yr$^{-1}$.  For instance, these rates
are a factor of $10-100$ times smaller than the \ha\ derived star formation
rates of $z\sim2$ SMGs, also uncorrected for reddening \citep[Figure \ref{fig:SFR};
SMG data from][]{Swinbank04}.  While the extinction corrections for DOGs should
be large --- thought to exceed a factor 70 in one of the sample members ---
even the reddening corrected star formation rates are relatively modest.  
The rates are estimated to be at the LIRG level rather than $100-1000$
\sm\ yr$^{-1}$ expected in extinction corrected star formation rates of ULIRGs and SMGs.

These low star formation rates may be understandable in a scenario where the DOGs are  observed in a post merger phase and star formation has begun to shut down.  \citet{Narayanan09} attempts to
simulate galaxies with the unusual selection criteria of DOGs by inducing
mergers in massive gas rich disks.  The \citet{Narayanan09} merger simulations
actually produce objects that would be selected as DOGs.  In the simulations,
AGN dominated DOGs only appear after peak star formation.  In fact, it is
precisely this AGN activity that allows these objects to remain DOGs even when
the star formation rate has dropped below the ULIRG level.

The observed kinematics of the DOGs can not rule out recent merging activity.
There is no evidence for well ordered rotation in extended star forming disks,
and the the narrow \ha\ lines show kinematic offsets from the broad \ha\ lines
of $150-500$ km s$^{-1}$.

It is more difficult to explain the colors and luminosities of the DOGs without
mergers, but some form of rapid gas accretion from cold flows might work
\citep[e.g.][]{Brooks09}. However, the metallicity of the star forming gas
appears to be metal rich with 12 + log(O/H) $\sim8.6$ (solar).  Thus, a
scenario of pristine gas infall is less likely than a scenario of a post merger
system where previous star formation has polluted the gas with metals.


\section{Conclusions}
We have obtained high spatial resolution IFU observations of four $z>1$ extremely dust obscured galaxies with the spectroscopy centered on the \ha\ emission line.   The \ha\ emission of all four DOGs are dominated by a broad-line component with velocity line widths in excess of 2000 km/s.  In contrast, the spatially offset narrow \ha\ lines have velocity dispersions $< 600$ km/s. 

We use the broad \ha\ line widths and flux measurements to estimate the BH masses of these systems.  After corrections for dust obscuration, the BHs of the DOGs are small for their host galaxy luminosity compared with $z\sim2$ and local unobscured AGN. Much of this offset can be explained if the DOGs have significant young stellar populations that fade rapidly.  Assuming fading from an instantaneous burst, the DOGs fade to the local BH-mass/host-luminosity relation by today.  Alternatively the dust correction we have applied to the DOG host luminosities could be high. Assuming no dust correction for the galaxy luminosity and a passive fading rate, the DOGs still evolve onto the local BH-mass/galaxy-luminosity relation.  

Interestingly, the $z\sim2$ AGN actually fade past the local relation even with only a passive fading applied.  Thus the comparison samples need to grow in stellar mass to reach the local relation today.  
If the DOGs are in a BH growth phase, they could easily reach the $z\sim2$ relation (after accounting for fading) by growing their BHs at an Eddington rate for $50-100$ Myrs.   If they were to do so, they too would need to grow in stellar mass to reach the local relation.

The SFRs of the DOGs are also surprising low; after correcting for dust obscuration the SFRs are still $<100$ M$_{\odot}$ yr$^{-1}$.  Thus, without the presence of an AGN these galaxies would be classified as LIRGs rather than ULIRGs.  A merger scenario in which star formation peaks prior to the AGN growth phase  could explain the observations.  The measured kinematics of the DOGs can not rule out a merger scenario. 
   
\acknowledgments
The adaptive optics data used in this study were obtained at the Keck Observatory, which is operated as a scientific partnership among Caltech, UC, and NASA.  Special thanks to the Keck support astronomer staff, especially Jim Lyke who worked on the OSIRIS calibration during the warm phase, when most of these data were taken.  Also, special thanks to Shelley Wright for providing guidance on the data reduction. The authors wish to recognize and acknowledge the very significant cultural role and reverence that the summit of Mauna Kea has always had within the indigenous Hawaiian community. The laser guide star adaptive optics system was funded by the W. M. Keck Foundation. 

\bibliographystyle{/Users/jmel/bib/apj}
\bibliography{/Users/jmel/bib/bigbib2}

\begin{thebibliography}{39}
\expandafter\ifx\csname natexlab\endcsname\relax\def\natexlab#1{#1}\fi

\bibitem[{{Bettoni} {et~al.}(2003){Bettoni}, {Falomo}, {Fasano}, \&
  {Govoni}}]{Bettoni03}
{Bettoni}, D., {Falomo}, R., {Fasano}, G., \& {Govoni}, F. 2003, \aap, 399, 869

\bibitem[{{Blain} {et~al.}(2004){Blain}, {Chapman}, {Smail}, \&
  {Ivison}}]{Blain04}
{Blain}, A.~W., {Chapman}, S.~C., {Smail}, I., \& {Ivison}, R. 2004, \apj, 611,
  725

\bibitem[{{Borys} {et~al.}(2005){Borys}, {Smail}, {Chapman}, {Blain},
  {Alexander}, \& {Ivison}}]{Borys05}
{Borys}, C., {Smail}, I., {Chapman}, S.~C., {Blain}, A.~W., {Alexander}, D.~M.,
  \& {Ivison}, R.~J. 2005, \apj, 635, 853

\bibitem[{{Brand} {et~al.}(2007){Brand}, {Dey}, {Desai}, {Soifer}, {Bian},
  {Armus}, {Brown}, {Le Floc'h}, {Higdon}, {Houck}, {Jannuzi}, \&
  {Weedman}}]{Brand07}
{Brand}, K., {Dey}, A., {Desai}, V., {Soifer}, B.~T., {Bian}, C., {Armus}, L.,
  {Brown}, M.~J.~I., {Le Floc'h}, E., {Higdon}, S.~J., {Houck}, J.~R.,
  {Jannuzi}, B.~T., \& {Weedman}, D.~W. 2007, \apj, 663, 204

\bibitem[{{Brodwin} {et~al.}(2008){Brodwin}, {Dey}, {Brown}, {Pope}, {Armus},
  {Bussmann}, {Desai}, {Jannuzi}, \& {Le Floc'h}}]{Brodwin08}
{Brodwin}, M., {Dey}, A., {Brown}, M.~J.~I., {Pope}, A., {Armus}, L.,
  {Bussmann}, S., {Desai}, V., {Jannuzi}, B.~T., \& {Le Floc'h}, E. 2008,
  \apjl, 687, L65

\bibitem[{{Brooks} {et~al.}(2009){Brooks}, {Governato}, {Quinn}, {Brook}, \&
  {Wadsley}}]{Brooks09}
{Brooks}, A.~M., {Governato}, F., {Quinn}, T., {Brook}, C.~B., \& {Wadsley}, J.
  2009, \apj, 694, 396

\bibitem[{{Bussmann} {et~al.}(2009){Bussmann}, {Dey}, {Lotz}, {Armus}, {Brand},
  {Brown}, {Desai}, {Eisenhardt}, {Higdon}, {Higdon}, {Jannuzi}, {LeFloc'h},
  {Melbourne}, {Soifer}, \& {Weedman}}]{Bussmann09}
{Bussmann}, R.~S., {Dey}, A., {Lotz}, J., {Armus}, L., {Brand}, K., {Brown},
  M.~J.~I., {Desai}, V., {Eisenhardt}, P., {Higdon}, J., {Higdon}, S.,
  {Jannuzi}, B.~T., {LeFloc'h}, E., {Melbourne}, J., {Soifer}, B.~T., \&
  {Weedman}, D. 2009, \apj, 693, 750

\bibitem[{{Cardelli} {et~al.}(1989){Cardelli}, {Clayton}, \&
  {Mathis}}]{Cardelli89}
{Cardelli}, J.~A., {Clayton}, G.~C., \& {Mathis}, J.~S. 1989, \apj, 345, 245

\bibitem[{{Chary} \& {Elbaz}(2001)}]{Chary01}
{Chary}, R., \& {Elbaz}, D. 2001, \apj, 556, 562

\bibitem[{{Desai} {et~al.}(2008){Desai}, {Soifer}, {Dey}, {Jannuzi}, {Le
  Floc'h}, {Bian}, {Brand}, {Brown}, {Armus}, {Weedman}, {Cool}, {Stern}, \&
  {Brodwin}}]{Desai08}
{Desai}, V., {Soifer}, B.~T., {Dey}, A., {Jannuzi}, B.~T., {Le Floc'h}, E.,
  {Bian}, C., {Brand}, K., {Brown}, M.~J.~I., {Armus}, L., {Weedman}, D.~W.,
  {Cool}, R., {Stern}, D., \& {Brodwin}, M. 2008, \apj, 679, 1204

\bibitem[{{Dey} {et~al.}(2008){Dey}, {Soifer}, {Desai}, {Brand}, {Le Floc'h},
  {Brown}, {Jannuzi}, {Armus}, {Bussmann}, {Brodwin}, {Bian}, {Eisenhardt},
  {Higdon}, {Weedman}, \& {Willner}}]{Dey08}
{Dey}, A., {Soifer}, B.~T., {Desai}, V., {Brand}, K., {Le Floc'h}, E., {Brown},
  M.~J.~I., {Jannuzi}, B.~T., {Armus}, L., {Bussmann}, S., {Brodwin}, M.,
  {Bian}, C., {Eisenhardt}, P., {Higdon}, S.~J., {Weedman}, D., \& {Willner},
  S.~P. 2008, \apj, 677, 943

\bibitem[{{Elias} {et~al.}(1982){Elias}, {Frogel}, {Matthews}, \&
  {Neugebauer}}]{Elias82}
{Elias}, J.~H., {Frogel}, J.~A., {Matthews}, K., \& {Neugebauer}, G. 1982, \aj,
  87, 1029

\bibitem[{{Ferrarese} \& {Ford}(2005)}]{Ferrarese05}
{Ferrarese}, L., \& {Ford}, H. 2005, \ssr, 116, 523

\bibitem[{{Greene} \& {Ho}(2005)}]{Greene05}
{Greene}, J.~E., \& {Ho}, L.~C. 2005, \apj, 630, 122

\bibitem[{{Houck} {et~al.}(2005){Houck}, {Soifer}, {Weedman}, {Higdon},
  {Higdon}, {Herter}, {Brown}, {Dey}, {Jannuzi}, {Le Floc'h}, {Rieke}, {Armus},
  {Charmandaris}, {Brandl}, \& {Teplitz}}]{Houck05}
{Houck}, J.~R., {Soifer}, B.~T., {Weedman}, D., {Higdon}, S.~J.~U., {Higdon},
  J.~L., {Herter}, T., {Brown}, M.~J.~I., {Dey}, A., {Jannuzi}, B.~T., {Le
  Floc'h}, E., {Rieke}, M., {Armus}, L., {Charmandaris}, V., {Brandl}, B.~R.,
  \& {Teplitz}, H.~I. 2005, \apjl, 622, L105

\bibitem[{{Jannuzi} \& {Dey}(1999)}]{JannuziDey99}
{Jannuzi}, B.~T., \& {Dey}, A. 1999, in Astronomical Society of the Pacific
  Conference Series, Vol. 191, Photometric Redshifts and the Detection of High
  Redshift Galaxies, ed. R.~{Weymann}, L.~{Storrie-Lombardi}, M.~{Sawicki}, \&
  R.~{Brunner}, 111--+

\bibitem[{{Kennicutt}(1998)}]{Kennicutt98}
{Kennicutt}, R.~C. 1998, \araa, 36, 189

\bibitem[{{Kormendy} \& {Gebhardt}(2001)}]{Kormendy01}
{Kormendy}, J., \& {Gebhardt}, K. 2001, in American Institute of Physics
  Conference Series, Vol. 586, 20th Texas Symposium on relativistic
  astrophysics, ed. {J.~C.~Wheeler \& H.~Martel}, 363--381

\bibitem[{{Kukula} {et~al.}(2001){Kukula}, {Dunlop}, {McLure}, {Miller},
  {Percival}, {Baum}, \& {O'Dea}}]{Kukula01}
{Kukula}, M.~J., {Dunlop}, J.~S., {McLure}, R.~J., {Miller}, L., {Percival},
  W.~J., {Baum}, S.~A., \& {O'Dea}, C.~P. 2001, \mnras, 326, 1533

\bibitem[{{Law} {et~al.}(2009){Law}, {Steidel}, {Erb}, {Larkin}, {Pettini},
  {Shapley}, \& {Wright}}]{Law09}
{Law}, D.~R., {Steidel}, C.~C., {Erb}, D.~K., {Larkin}, J.~E., {Pettini}, M.,
  {Shapley}, A.~E., \& {Wright}, S.~A. 2009, \apj, 697, 2057

\bibitem[{{Magliocchetti} {et~al.}(2007){Magliocchetti}, {Silva}, {Lapi}, {de
  Zotti}, {Granato}, {Fadda}, \& {Danese}}]{Magliocchetti07}
{Magliocchetti}, M., {Silva}, L., {Lapi}, A., {de Zotti}, G., {Granato}, G.~L.,
  {Fadda}, D., \& {Danese}, L. 2007, \mnras, 375, 1121

\bibitem[{{Maiolino} {et~al.}(2007){Maiolino}, {Shemmer}, {Imanishi}, {Netzer},
  {Oliva}, {Lutz}, \& {Sturm}}]{Maiolino07}
{Maiolino}, R., {Shemmer}, O., {Imanishi}, M., {Netzer}, H., {Oliva}, E.,
  {Lutz}, D., \& {Sturm}, E. 2007, \aap, 468, 979

\bibitem[{{Melbourne} {et~al.}(2008{\natexlab{a}}){Melbourne}, {Ammons},
  {Wright}, {Metevier}, {Steinbring}, {Max}, {Koo}, {Larkin}, \&
  {Barczys}}]{Melbourne08a}
{Melbourne}, J., {Ammons}, M., {Wright}, S.~A., {Metevier}, A., {Steinbring},
  E., {Max}, C., {Koo}, D.~C., {Larkin}, J.~E., \& {Barczys}, M.
  2008{\natexlab{a}}, \aj, 135, 1207

\bibitem[{{Melbourne} {et~al.}(2009){Melbourne}, {Bussman}, {Brand}, {Desai},
  {Armus}, {Dey}, {Jannuzi}, {Houck}, {Matthews}, \& {Soifer}}]{Melbourne09}
{Melbourne}, J., {Bussman}, R.~S., {Brand}, K., {Desai}, V., {Armus}, L.,
  {Dey}, A., {Jannuzi}, B.~T., {Houck}, J.~R., {Matthews}, K., \& {Soifer},
  B.~T. 2009, \aj, 137, 4854

\bibitem[{{Melbourne} {et~al.}(2008{\natexlab{b}}){Melbourne}, {Desai},
  {Armus}, {Dey}, {Brand}, {Thompson}, {Soifer}, {Matthews}, {Jannuzi}, \&
  {Houck}}]{Melbourne08b}
{Melbourne}, J., {Desai}, V., {Armus}, L., {Dey}, A., {Brand}, K., {Thompson},
  D., {Soifer}, B.~T., {Matthews}, K., {Jannuzi}, B.~T., \& {Houck}, J.~R.
  2008{\natexlab{b}}, \aj, 136, 1110

\bibitem[{{Melbourne} \& {Salzer}(2002)}]{Melbourne02}
{Melbourne}, J., \& {Salzer}, J.~J. 2002, \aj, 123, 2302

\bibitem[{{Narayanan} {et~al.}(2009{\natexlab{a}}){Narayanan}, {Dey},
  {Hayward}, {Cox}, {Bussmann}, {Brodwin}, {Jonsson}, {Hopkins}, {Groves},
  {Younger}, \& {Hernquist}}]{Narayanan09a}
{Narayanan}, D., {Dey}, A., {Hayward}, C., {Cox}, T.~J., {Bussmann}, R.~S.,
  {Brodwin}, M., {Jonsson}, P., {Hopkins}, P., {Groves}, B., {Younger}, J.~D.,
  \& {Hernquist}, L. 2009{\natexlab{a}}, ArXiv e-prints

\bibitem[{{Narayanan} {et~al.}(2009{\natexlab{b}}){Narayanan}, {Hayward},
  {Cox}, {Hernquist}, {Jonsson}, {Younger}, \& {Groves}}]{Narayanan09}
{Narayanan}, D., {Hayward}, C.~C., {Cox}, T.~J., {Hernquist}, L., {Jonsson},
  P., {Younger}, J.~D., \& {Groves}, B. 2009{\natexlab{b}}, ArXiv e-prints

\bibitem[{{Osterbrock}(1989)}]{Osterbrock89}
{Osterbrock}, D.~E. 1989, {Astrophysics of gaseous nebulae and active galactic
  nuclei}, ed. {Osterbrock, D.~E.}

\bibitem[{{Peng} {et~al.}(2006{\natexlab{a}}){Peng}, {Impey}, {Ho}, {Barton},
  \& {Rix}}]{Peng06}
{Peng}, C.~Y., {Impey}, C.~D., {Ho}, L.~C., {Barton}, E.~J., \& {Rix}, H.
  2006{\natexlab{a}}, \apj, 640, 114

\bibitem[{{Peng} {et~al.}(2006{\natexlab{b}}){Peng}, {Impey}, {Rix},
  {Kochanek}, {Keeton}, {Falco}, {Leh{\'a}r}, \& {McLeod}}]{Peng06a}
{Peng}, C.~Y., {Impey}, C.~D., {Rix}, H., {Kochanek}, C.~S., {Keeton}, C.~R.,
  {Falco}, E.~E., {Leh{\'a}r}, J., \& {McLeod}, B.~A. 2006{\natexlab{b}}, \apj,
  649, 616

\bibitem[{{Ridgway} {et~al.}(2001){Ridgway}, {Heckman}, {Calzetti}, \&
  {Lehnert}}]{Ridgway01}
{Ridgway}, S.~E., {Heckman}, T.~M., {Calzetti}, D., \& {Lehnert}, M. 2001,
  \apj, 550, 122

\bibitem[{{Rothberg} \& {Joseph}(2006)}]{Rothberg06}
{Rothberg}, B., \& {Joseph}, R.~D. 2006, \aj, 131, 185

\bibitem[{{Swinbank} {et~al.}(2004){Swinbank}, {Smail}, {Chapman}, {Blain},
  {Ivison}, \& {Keel}}]{Swinbank04}
{Swinbank}, A.~M., {Smail}, I., {Chapman}, S.~C., {Blain}, A.~W., {Ivison},
  R.~J., \& {Keel}, W.~C. 2004, \apj, 617, 64

\bibitem[{{van Dokkum} \& {Franx}(2001)}]{van-Dokkum01}
{van Dokkum}, P.~G., \& {Franx}, M. 2001, \apj, 553, 90

\bibitem[{{Vestergaard}(2002)}]{Vestergaard02}
{Vestergaard}, M. 2002, \apj, 571, 733

\bibitem[{{Wright} {et~al.}(2007){Wright}, {Larkin}, {Barczys}, {Erb},
  {Iserlohe}, {Krabbe}, {Law}, {McElwain}, {Quirrenbach}, {Steidel}, \&
  {Weiss}}]{Wright07}
{Wright}, S.~A., {Larkin}, J.~E., {Barczys}, M., {Erb}, D.~K., {Iserlohe}, C.,
  {Krabbe}, A., {Law}, D.~R., {McElwain}, M.~W., {Quirrenbach}, A., {Steidel},
  C.~C., \& {Weiss}, J. 2007, \apj, 658, 78

\bibitem[{{Wright} {et~al.}(2010){Wright}, {Larkin}, {Graham}, \&
  {Ma}}]{Wright10}
{Wright}, S.~A., {Larkin}, J.~E., {Graham}, J.~R., \& {Ma}, C. 2010, \apj, 711,
  1291

\bibitem[{{Wright} {et~al.}(2009){Wright}, {Larkin}, {Law}, {Steidel},
  {Shapley}, \& {Erb}}]{Wright09}
{Wright}, S.~A., {Larkin}, J.~E., {Law}, D.~R., {Steidel}, C.~C., {Shapley},
  A.~E., \& {Erb}, D.~K. 2009, \apj, 699, 421

\end{thebibliography}
\clearpage

\end{document}